# Role of transparency of platinum-ferromagnet interface in determining intrinsic magnitude of spin Hall effect

Weifeng Zhang[1,2*], Wei Han[1,3,4*], Xin Jiang[1†], See-Hun Yang[1] and Stuart S. P. Parkin[1,5§]

[1]IBM Almaden Research Center, San Jose, California 95120, USA

[2]Department of Material Science Engineering, Stanford University, Stanford, California 94305, USA

[3]International Center for Quantum Materials, School of Physics, Peking University, Beijing 100871, China

[4]Collaborative Innovation Center of Quantum Matter, Beijing 100871, P. R. China

[5]Max Planck Institute for Microstructure Physics, Halle (Saale), D-006120 Germany

*These authors contributed equally to the work

[†] Present address: Western Digital Corporation, Fremont, California 94539, USA

[§] Email: stuart.parkin@mpi-halle.mpg.de

**The spin Hall effect (SHE) converts charge current to pure spin currents in orthogonal directions in materials that have significant spin-orbit coupling. The efficiency of the conversion is described by the spin Hall Angle (SHA). The SHA can most readily be inferred by using the generated spin currents to excite or rotate the magnetization of ferromagnetic films or nano-elements via spin-transfer torques. Some of the largest spin torque derived spin Hall angles (ST-SHA) have been reported in platinum. Here we show,**

**using spin torque ferromagnetic resonance (ST-FMR) measurements, that the transparency of the Pt-ferromagnet interface to the spin current plays a central role in determining the magnitude of the ST-SHA. We measure a much larger ST-SHA in Pt/cobalt (~0.11) compared to Pt/permalloy (~0.05) bilayers when the interfaces are assumed to be completely transparent. Taking into account the transparency of these interfaces, as derived from spin–mixing conductances, we find that the intrinsic SHA in platinum has a much higher value of 0.19 ± 0.04 as compared to the ST-SHA. The importance of the interface transparency is further exemplified by the insertion of atomically thin magnetic layers at the Pt/permalloy interface that we show strongly modulates the magnitude of the ST-SHA.**

The spin Hall effect (SHE), namely the generation of pure spin current density, $\hbar/2 J_S$, from charge current density, $J_C$, via spin-orbit interactions[1-3], has significant potential for next generation spintronic devices[4]. For example, it has been demonstrated that spin currents derived from SHE can be used to switch the magnetization of ferromagnets (FMs)[5,6] and to move domain walls very efficiently via a chiral spin torque mechanism in perpendicularly magnetized ferromagnetic layers[7-10]. The magnitude of the SHE has often been derived from such a manipulation of the magnetization of a FM, via, for example, spin-torque ferromagnetic resonance (ST-FMR)[11-13] or spin-torque switching of perpendicularly magnetized films[5,6,14]. Both these measurements rely on spin-transfer torque derived from spin currents, generated in non-magnetic (NM) layers via SHE, diffusing into and thereby acting on the adjacent FMs. Here, we show that the transparency of the NM-FM interface to the spin current plays a central role in determining the intrinsic value of the SHE in the NM metal and that reported SHE values are a lower bound.

The SHE of Pt in Pt/Py (Py=permalloy) and Pt/Co bilayer film structures is characterized in a ST-FMR experiment, as shown schematically in Fig. 1a. The films are deposited by magnetron sputtering and devices are fabricated from the films by optical lithography and argon ion beam etching (see Methods for details). The devices are in the form of rectangular strips that are 100 μm long and 10 μm wide. An external magnetic field $\vec{H}_{ext}$ is applied at 45° to the length of the strip, and a RF microwave current is applied to the device via the AC port of a bias tee (Fig. 1a). The magnetization of the FM layer is affected by the RF current flowing in the Pt layer from two torques, namely that from the oscillating magnetic field, and that from the oscillating spin current. For a FM layer with magnetization $\hat{m}$, the field torque ($\tau_H$ in Fig. 1a) is equal to $\gamma \hat{m} \times \vec{H}_{rf}$, in which $\gamma$ is the gyromagnetic ratio and $\vec{H}_{rf}$ is the Oersted field generated by the RF current in the Pt layer. The spin Hall torque ($\tau_{ST}$ in Fig. 1a) is expressed as $\gamma \frac{\hbar}{2\mu_0 M_S t} J_{SH}^{ST} (\hat{m} \times \hat{\sigma} \times \hat{m})$, in which $\hbar/2 J_{SH}^{ST}$ is the spin current density generated via SHE in the Pt layer that diffuses into the FM layer, $e$ is the electron charge, $\mu_0$ is the permeability in vacuum, $M_S$ is the saturation magnetization of the FM layer, $t$ is the thickness of the magnetic layer, and $\hat{\sigma}$ is the direction of the injected spin moment. The time derivative of the magnetization of the FM layer can then be expressed by the Landau–Lifshitz–Gilbert equation, as follows[11,15]:

$$\frac{d\hat{m}}{dt} = -\gamma \hat{m} \times \vec{H}_{eff} + \alpha \hat{m} \times \frac{d\hat{m}}{dt} + \gamma \frac{\hbar}{2\mu_0 M_S t} J_{SH}^{ST} (\hat{m} \times \hat{\sigma} \times \hat{m}) - \gamma \hat{m} \times \vec{H}_{rf}$$

where $\hbar$ is the reduced Planck's constant, $\vec{H}_{eff}$ is the sum of $\vec{H}_{ext}$ and the out-of-plane demagnetization field, and $\alpha$ is the Gilbert damping constant.

Figs. 1b and 1c show ST-FMR spectra measured on 60 Pt/55 Py and 60 Pt/52 Co bilayers (the numbers are the layer thicknesses in angstrom) for RF frequencies varying between 9 and 13 GHz. $V_{mix}$ is plotted versus $H_{ext}$ for various RF frequencies $f$. $V_{mix}$ is the DC voltage generated across the device that arises from mixing of the RF charge current through the device and the consequent oscillating magnetization of the FM layer that affects the device resistance through its magnetoresistance. As shown in Figs. 1b and 1c, $V_{mix}$ shows significant values only under resonant conditions where the magnitude of $\vec{H}_{ext} \sim H_{res}$, as discussed later.

The spin Hall angle (SHA) is the ratio of the spin current density to the RF current density in the Pt layer of thickness $d$, and is given by[11]:

$$\theta_{SH}^{ST} = \frac{eJ_{SH}^{ST}}{J_c} = \frac{S}{A} \frac{e\mu_0 M_S t d}{\hbar} [1+(\frac{4\pi M_{eff}}{H_{res}})]^{\frac{1}{2}}$$

where $S = \hbar \frac{J_{SH}^{ST}}{2\mu_0 M_S t}$, $A = H_{rf}[1+(4\pi M_{eff}/H_{res})]^{\frac{1}{2}}$, and $M_{eff}$ is the effective magnetization that depends on both $M_S$ and the out-of-plane anisotropy field (see S2). The values of $S$ and $A$, the resonance field ($H_{res}$) and the half line-width ($\Delta H$) are obtained by fitting $V_{mix}$ with symmetric and anti-symmetric Lorentzian functions, according to:

$$V_{mix} = c[S\frac{\Delta H^2}{\Delta H^2 + (H_{ext} - H_{res})^2} + A\frac{\Delta H(H_{ext} - H_{res})}{\Delta H^2 + (H_{ext} - H_{res})^2}] \quad (1)$$

where $c$ is a constant. Typical experimental curves and the fits to these curves using equation (1) are shown in the insets to Figs. 1d and 1e, for 60 Pt/55 Py and 60 Pt/52 Co, respectively, for $f = 9$ GHz. $M_{eff}$ is extracted by fitting the frequency $f_{res}$ as a function of the resonant field using the Kittel formula[16] $f_{res} = (\frac{\gamma}{2\pi})[H_{res}(H_{res} + 4\pi M_{eff})]^{\frac{1}{2}}$. $M_S$ is measured using vibrating sample magnetometry on the un-patterned film prior to device fabrication. $M_{eff}$ is similar in value to $M_S$ when $t \geq 50$ Å: for smaller thicknesses an interface perpendicular magnetic anisotropy leads to a reduced value of $M_{eff}$ (see Figs. S2 and S3).

The values of $\theta_{SH}^{ST}$ derived from the ST-FMR measurements for Pt in 60 Pt/55 Py and 60 Pt/52 Co bilayers are plotted in Figs. 1d and 1e as a function of the RF frequency. The SHA values do not exhibit any significant dependence on the RF frequency but the value of the SHA of Pt in Pt/Py is 0.05 ± 0.01 while the value for Pt in Pt/Co is about two times higher at 0.11 ± 0.02. The error bars (calculated for a 95% confidence value) are from values of $S/A$ that were averaged on from ~ 5 to 15 devices. We have carefully estimated the role of potential artifacts in our ST-FMR experiments, including inhomogeneous Oersted field and current distributions, phase shifts, spin-pumping, and field-like versus damping torques. We conclude that these effects play a minor role in our experiments (see supplementary information 4-6). Particularly, the DC voltage due to spin pumping is estimated to be much smaller than the magnitude of the symmetric component of the ST-FMR voltage signal (<1% for Pt/Py and ~5% for Pt/Co).

The thicknesses of the Co and Py layers are varied to see whether these influence the determination of the SHA in Pt. Fig. 2 shows FMR spectra at 9 GHz for (a) 60Pt/$t$ Py, for $t$ = 33, 55, and 110 Å, and, (b) 60Pt/$t$ Co at 9 GHz, for $t$ = 39, 52, and 65 Å. For thinner films, the magnetoresistance of the Co and Py is too small to give a large enough $V_{mix}$ to obtain reliable values of $S/A$. As shown in Figs. 2c and 2d, $\theta_{SH}^{ST}$ exhibits a weak dependence on the thicknesses of the Py or Co layers, even though, as shown in Figs. 2e and 2f, $M_{eff}$ and $\alpha$ vary significantly as a function of the corresponding FM layer thickness.

From the analysis of the ST-FMR data, it is clear that the derived value of $\theta_{SH}^{ST}$ for Pt in the Pt/Co bilayer structure is much higher than that for Pt in the Pt/Py bilayer. To understand this very significant difference in the measured values of $\theta_{SH}^{ST}$ for Pt using Co and Py layers, we postulate that this difference relates to the transparency, $T$, of the interface, i.e. the spin current density that diffuses into the FM layer, $\hbar/2 J_{SH}^{ST}$, is smaller than the actual spin current density generated via the SHE in the Pt layer, $\hbar/2 J_{SH}$. To calculate $T$ we first use a model[17] that was developed to account for the spin Hall magnetoresistance effect to relate $J_{SH}^{ST}$ to $J_{SH}$. This model relates $T$ to the spin-mixing conductance[18], $G_{\uparrow\downarrow}$, at the Pt/FM interface via the relationship:

$$T = \frac{G_{\uparrow\downarrow} \tanh(\frac{d}{2\lambda})}{G_{\uparrow\downarrow} \coth(\frac{d}{\lambda}) + \frac{\sigma_{Pt}}{\lambda} \frac{h}{2e^2}} \quad (2)$$

where $d$ is the thickness of the Pt layer (see Fig. 3a and 3b), $\lambda$ and $\sigma_{Pt}$ are the spin diffusion length and the conductivity of Pt, respectively, and $h$ is Plank's constant. (Note that we make the assumption that the imaginary part of $G_{\uparrow\downarrow}$ is much smaller than its real part[19,20]). When $\lambda$ is significantly smaller than $d$, $G_{\uparrow\downarrow}$ depends on the effective spin-mixing conductance, $G_{eff}$, according to the relationship[21,22]:

$$G_{\uparrow\downarrow} = G_{eff} \frac{\frac{\sigma_{Pt}}{\lambda} \frac{h}{2e^2}}{\frac{\sigma_{Pt}}{\lambda} \frac{h}{2e^2} - G_{eff}} \quad (3)$$

where $G_{eff}$ is given by[23,24]:

$$G_{eff} = \frac{4\pi M_S t}{g\mu_B}(\alpha_{Pt/FM} - \alpha_{FM}) \qquad (4)$$

Equations (3) and (4) are derived from theoretical models of spin pumping[20,25] that have been developed to account for the increased damping of a ferromagnetic layer when adjacent to a non-magnetic metallic layer due to spin currents that flow into and out of the FM layer into the non-magnetic layer when the magnetization of the FM layer is excited. Here we obtain the damping parameters of Pt/Py and Pt/Co bilayers, $\alpha_{Pt/FM}$, and those of FM layers of the same thickness without any Pt underlayers, $\alpha_{FM}$, from conventional FMR measurements using a strip-line technique[26]. We find that $(\alpha_{Pt/FM} - \alpha_{FM})M_S$ varies as the inverse FM layer thickness, as shown in Fig. 3c. From the slope of these curves we obtain $G_{eff}$ = (3.96±0.39)×10$^{19}$ m$^{-2}$ and (1.52±0.34)×10$^{19}$ m$^{-2}$, for Pt/Co and Pt/Py, respectively, using equation (4). These values quantitatively agree with previous reports of spin mixing conductances at Pt/FM metal interfaces[23,24,27]. From these values and those of the resistivity (15±1 μΩcm) and the spin diffusion length in Pt (14±2 Å) obtained from the dependence of the ST-SHA on the Pt layer thickness (Fig. S5), T is calculated to be 0.65 ± 0.06 and 0.25 ± 0.05 for Pt/Co and Pt/Py respectively, using equation (2).

From these experiments we find that the transparency of the Pt/Co interface is much higher than that of Pt/Py, which therefore can account for the much larger value of for Pt/Co compared to Pt/Py. Normalizing $\theta_{SH}^{ST}$ by T we obtain the intrinsic value of SHA for Pt, $\theta_{SH}$, which we find from Pt/Co is 0.17 ± 0.02, whereas for Pt/Py is 0.20 ± 0.03. Averaging these values we conclude that $\theta_{SH}$ = 0.19 ± 0.04 for Pt, a value which is higher than any experimental values reported to date for Pt.

To further test the role of interface transparency, $\theta_{SH}^{ST}$, magnetization, damping constant, resistivity and anisotropic magnetoresistance (AMR) are obtained for several $Co_{1-x}Ni_x$ alloys ($x$ = 0, 0.2, 0.4, 0.5, 0.6, 0.8, 1), as shown in Fig. 4. As the Co concentration is increased, both $\theta_{SH}^{ST}$ (Fig. 4a) and the spin-mixing conductances (Fig. 4b) at the Pt/$Co_{1-x}Ni_x$ interface increase, indicating the strong correlation between these parameters. The magnetization in the bilayer structure decreases as the Ni concentration increases (as shown in Fig. 4c), due to the larger magnetic moment of Co compared to Ni, following the Slater-Pauling curve[28]. The damping constant increases as the Ni concentration increases due to the larger damping of Ni than Co (Fig. 4d). These values are consistent with previous experimental and theoretical studies[29,30]. The resistivity and AMR, measured on single layer 500 Å thick $Co_{1-x}Ni_x$ films, are plotted in Fig. 4e versus $x$. These are consistent with previous studies[31]. There is no correlation between the ST-SHA and the AMR.

Finally, an ultrathin layer of thickness, δ = 0-16 Å, of various FM metals (Co, Fe, $Co_{0.5}Fe_{0.5}$ and Ni) is used to engineer the transparency of the spin Hall current in the Pt/Py bilayer system. The insertion layers are strongly exchange coupled to the Py layer and behave as a single magnetic unit (see, for example, [32]). As shown in Figs. 5a-5d, the ST- SHA highly depends on the composition and thickness of the inserted FM layer. For Co interface layers, $\theta_{SH}^{ST}$ increases and approaches the value of Co even for interface layers only ~ 4 Å thick, whereas for Ni interface layers $\theta_{SH}^{ST}$ is perhaps, not surprisingly, not much changed from that of Py ($Ni_{81}Fe_{19}$). For Fe and $Co_{0.5}Fe_{0.5}$, $\theta_{SH}^{ST}$ increases more slowly with interface thickness, perhaps due to the growth morphology of these layers. The interface transparency can be enhanced by matching the electronic properties of the NM and FM, reminiscent of the interfacial origin of the giant magnetoresistance effect[33]. These results further demonstrate the importance of the interface structure to the determination of the SHA.

The interface transparency effect that we have proposed accounts for the difference in the measured values of SHA for platinum is an effect distinct from "spin memory loss" that has been proposed to affect the magnitude of spin-pumping into platinum[34,35]. Spin memory loss is the loss of spin information due to spin-flip scattering at the interface. The spin angular momentum carried by the spin current is not transferred to the magnetic layer but presumably is transferred to the lattice through interfacial spin-orbit scattering. The interface transparency that we consider is an electronic effect whereby the transmission of the conduction electrons depends on the matching of the electronic bands in the two metals on either side of the interface, without any loss of spin polarization at the interface.

To unambiguously rule out a "spin memory loss" effect, we have introduced a Cu spacer layer at the Pt/Co and Pt/Py interfaces. Cu is a 3d element with a much smaller spin-orbit coupling parameter than Pt so the interfaces with Cu should have little spin-flip scattering, and, moreover, for the same reason the spin diffusion length in Cu is known to be very long. Indeed, what we find for Pt/Cu/Co is that there is little change in the magnitude of the SHA as the Cu spacer layer thickness is increased from zero to ~20 Å (Fig. 5e), with a small (~10%) decrease in the SHA for thin Cu layers. This small decrease could be due to the elimination of the proximity induced magnetization in the Pt layer[36] which depends on an interfacial exchange between the Co and Pt. The fact that the SHA does not significantly decrease with the introduction of a Cu layer can readily be rationalized in our interface transparency argument since it is well known from work on giant magnetoresistance that the Cu/Co interface is nearly transparent[37]. Thus, since we anticipate that the transparency of the Pt/Co interface will be similar to that of Pt/Cu due to the similar crystal and electronic structures, and since the Co/Cu interface is nearly transparent, the overall transparency is not much affected by a Cu spacer layer. On the other hand, for Pt/Cu/Py the Cu/Py interface is much less transparent than Cu/Co (i.e. the GMR is known to be much lower for Cu/Py than for Cu/Co[33]) so that the introduction of a Cu layer at the Pt/Py interface leads to a lower transparency and thus a reduced SHA (Fig. 5f). Moreover, when we calculate the spin Hall effect in Pt/Co and Pt/Py bilayers using the theory of spin memory loss[34,35], we find that the spin Hall angle in Pt/Co is predicted to be ~10 times larger than that in Pt/Py, which is not at all consistent with our experiments (see supplementary information).

Our results anticipate that spin currents generated from the SHE can be used more efficiently for applications such as manipulating series of domain walls in racetrack memories or for the excitation or switching of magnetic memory or logic elements by engineering the interface across which the spin current diffuses to allow for greater transparency. This may be accomplished by electronic band matching or by the use of insertion layers or graded interfaces.

**Methods**

**Films growth**

The films are grown on thermally oxidized Si substrates in a magnetron sputtering system with a base pressure of ~ $1 \times 10^{-8}$ Torr. The Pt and FM layers are grown at room temperature at a growth rate of ~1 Å/s. A 20 Å TaN capping layer is deposited *in situ* to avoid any oxidation of the FM layer. Since the resistivity of TaN is more than an order magnitude higher than either Pt or any of the FM layers used here, any current shunting through the capping layer is negligible.

**Device fabrication**

Photolithograpy and Argon ion etching are first used to form rectangular micro-strips (100 μm long and 10 μm wide). Then photo-lithography and ion beam deposition are used to fabricate two large Ru/gold pads (5 nm/50 nm thick) for electrical contacts. An optical image of a typical device (Pt/Py) is shown in the inset to Fig. 1b. The electrical measurements are performed using high-frequency probes.

**Device measurement**

An Agilent HP 83620B is used to provide the RF current for both the ST-FMR and conventional FMR measurements. For the ST-FMR studies the DC voltage is measured with a Keithley 2002 multimeter. An external magnetic field is applied at 45° to the microstrip for best sensitivity.

**Contributions**

W. Z., W. H., X. J. and S.S.P.P. designed the experiments. W.Z. performed the fabrication and measurements of the devices with help of W. H.. W. Z., W. H. and X. J. analyzed the data. X.J., W.H. and S.H. Y. grew the films. S.S.P.P. proposed and supervised the studies. W.Z, W.H. and S.S.P.P. wrote the manuscript.

**Acknowledgements**

We gratefully acknowledge help from Chris Lada in designing the ST-FMR measurement system and discussions with Tim Phung. W. H. also acknowledges support from the 1000

**Figure Captions**

**Figure 1 | ST-FMR measurement of the SHA of Pt in Pt/Py and Pt/Co. a,** Illustration of ST-FMR experiment. $H_{ext}$ is the external field. $\hat{m}$ is the magnetization. $\tau_H$ is the torque due to the Oersted field created by the RF charge current in Pt. $\tau_{ST}$ is the torque due to pure spin current generated by SHE. The RF power varies from 10 to 14 dBm and the RF frequency varies from 9 to 13 GHz. **b.** ST-FMR spectra measured on a 60 Pt/55 Py sample versus $H_{ext}$. Inset: optical image of a typical device. **c.** ST-FMR spectra measured on a 60 Pt/52 Co sample versus $H_{ext}$. Solid lines in **b,c** represent fits. **d,e,** Frequency dependence of measured SHA for 60 Pt/55 Py and 60 Pt/52 Co, respectively. Inset: FMR spectra measured for 60 Pt/55 Py and 60 Pt/52 Co at 9 GHz, respectively. The black solid lines are fits to the data using equation (3) where the green and red lines are the symmetric and anti-symmetric Lorentzian fits. The error bars (calculated for a 95% confidence value) are from values of S/A that were averaged on 5 - 15 devices.

**Figure 2 | Thickness dependence of ST-SHA for 60 Pt/*t* Py and 60 Pt/*t* Co. a,b,** ST-FMR spectra for various thicknesses of Py and Co at 9 GHz. Inset: schematic diagram of film structure. **c,d,** ST-SHA as a function of FM thickness, *t*. The error bars (calculated for a 95% confidence value) are from values of S/A that were averaged on 5 - 15 devices. **e,f,** Magnetization, $M_{eff}$ and damping constant, α, for Pt/Py and Pt/Co as a function of *t*.

**Figure 3 | Interface transparency of Pt/FM. a,b,** Schematic diagram showing interface transparency for Pt/Py and Pt/Co: the red and blue arrows represent the up and down spin accumulation. The white arrows show the electrons diffusing across or being reflected at the

interface. **c,** $(\alpha_{Pt/FM} - \alpha_{FM})M_S$ versus $1/t$. The dashed lines are linear fits to the data. Error bars correspond to one standard deviation.

**Figure 4 | ST- SHA and spin-mixing conductances of Pt/Co$_{1-x}$Ni$_x$. a, b,** ST-SHA and spin-mixing conductances measured in Pt/CoNi bilayer films for various Co$_{1-x}$Ni$_x$ compounds. The thicknesses of the Co$_{1-x}$Ni$_x$ layers are each ~60 Å. In **a** the error bars (calculated for a 95% confidence value) are from values of S/A that were averaged on 5 - 15 devices. In **b** error bars correspond to one standard deviation. **c,d,** Effective magnetization and damping constant obtained from ST-FMR measurements. **e,** Resistivity and AMR values measured on 500 Å thick Co$_{1-x}$Ni$_x$ calibration films.

**Figure 5 | Interface engineering of ST-SHA. a-d,** ST-SHA as a function of the thickness of magnetic insertion layers, δ, formed from Co, Fe, Co$_{0.5}$Fe$_{0.5}$, and Ni, inserted at the Pt/Py. **e-f,** ST-SHA as a function of the thickness of a Cu spacer layer, δ, inserted at the Pt/Co and Pt/Py interfaces. The error bars (calculated for a 95% confidence value) are from values of S/A that were averaged on 5 - 15 devices.

Figure 1

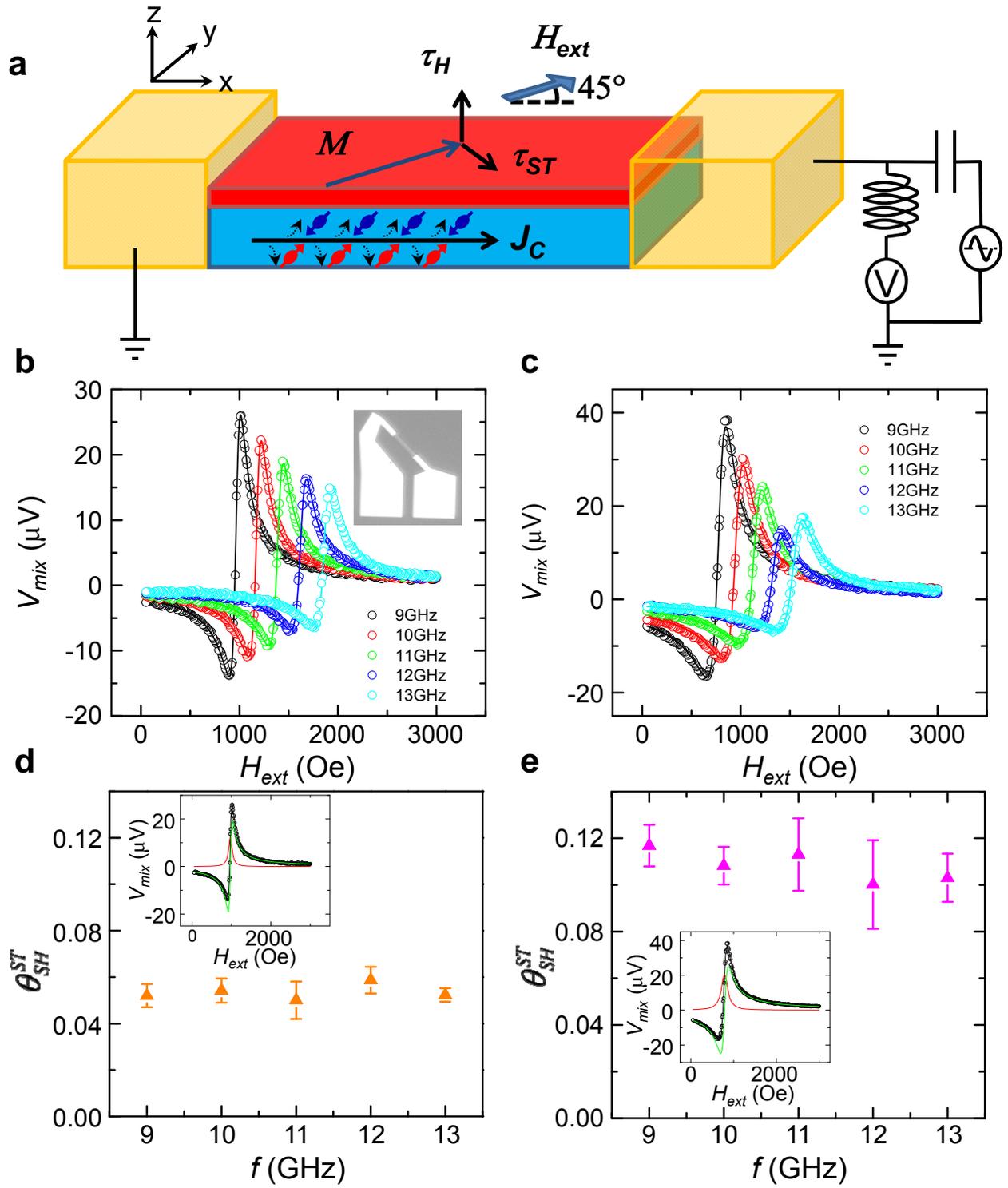

Figure 2

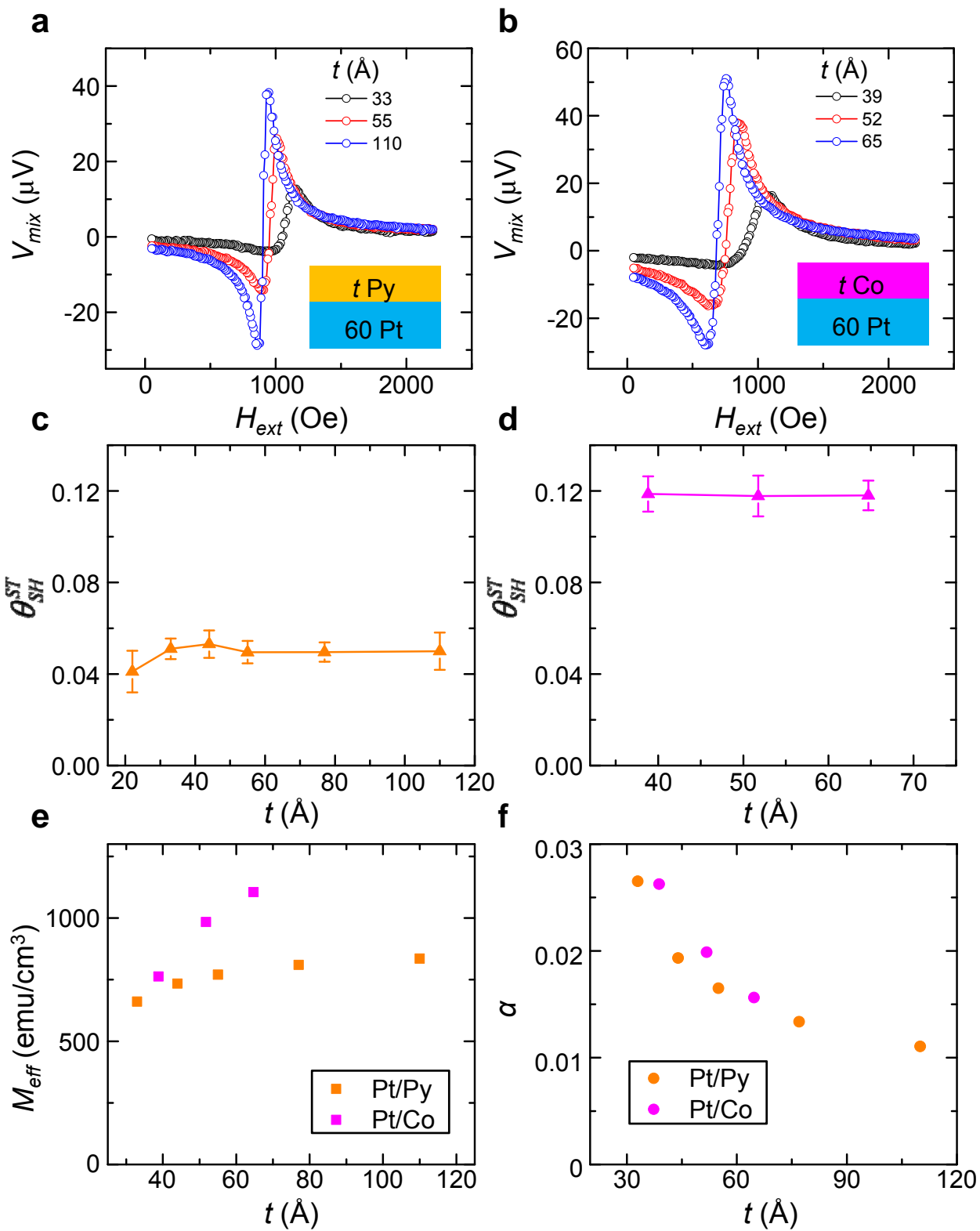

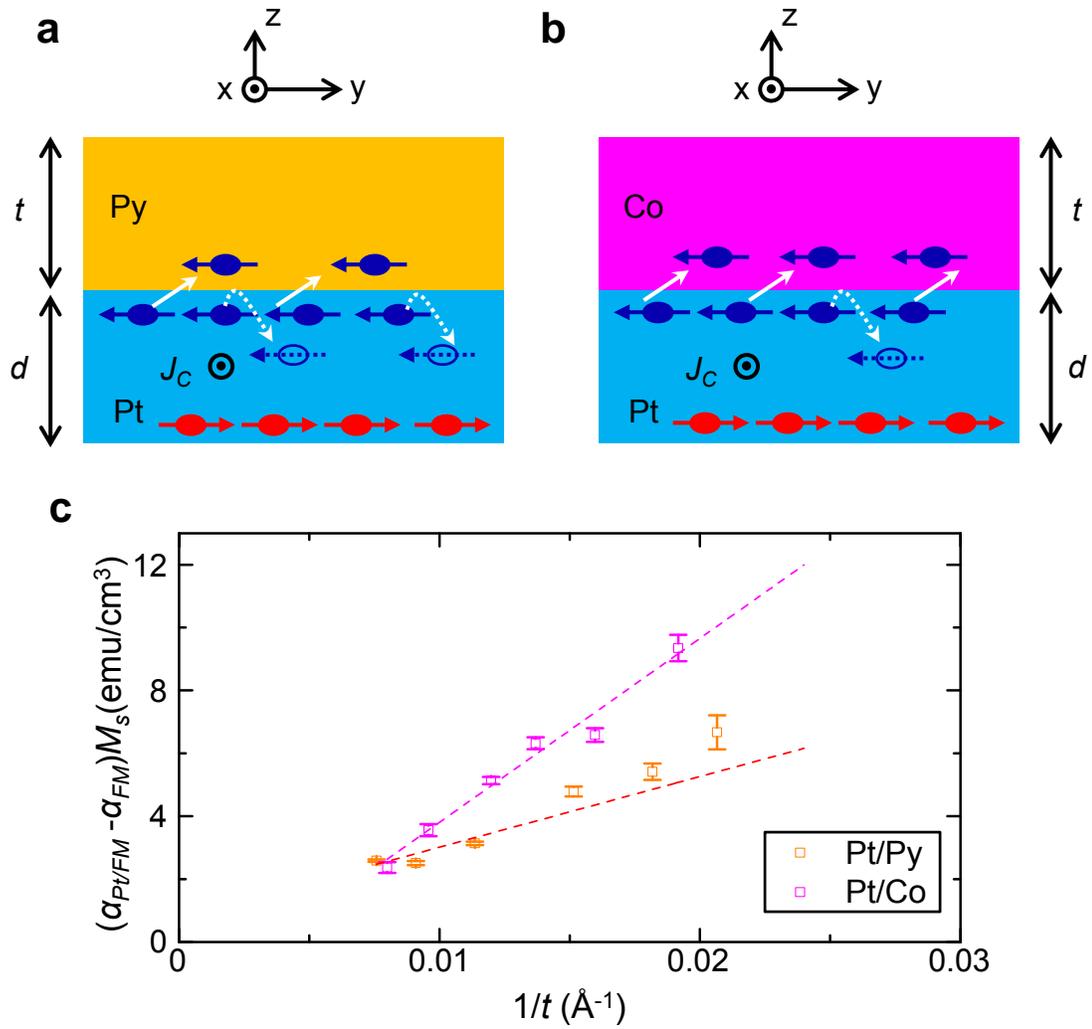

Figure 4

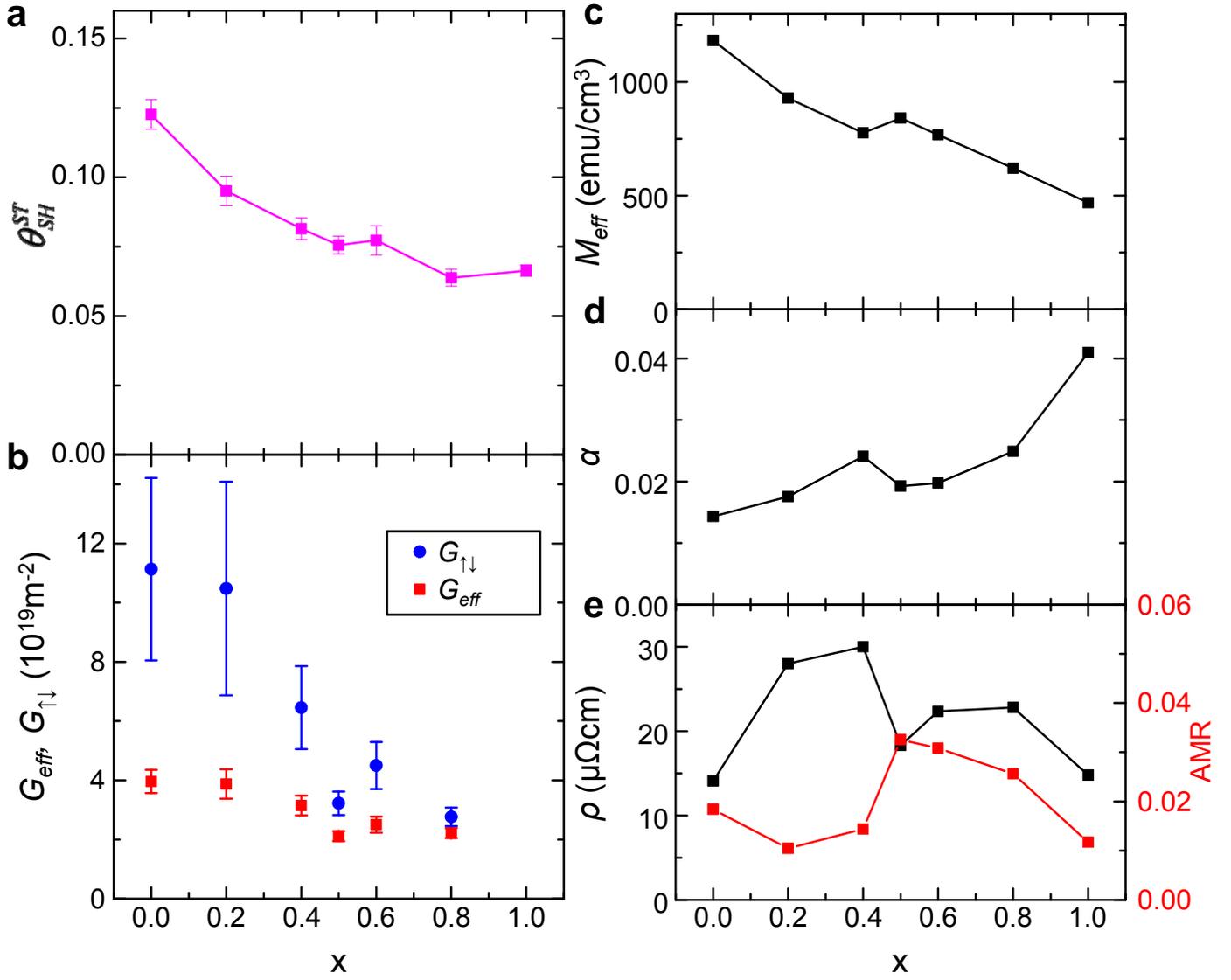

Figure 5

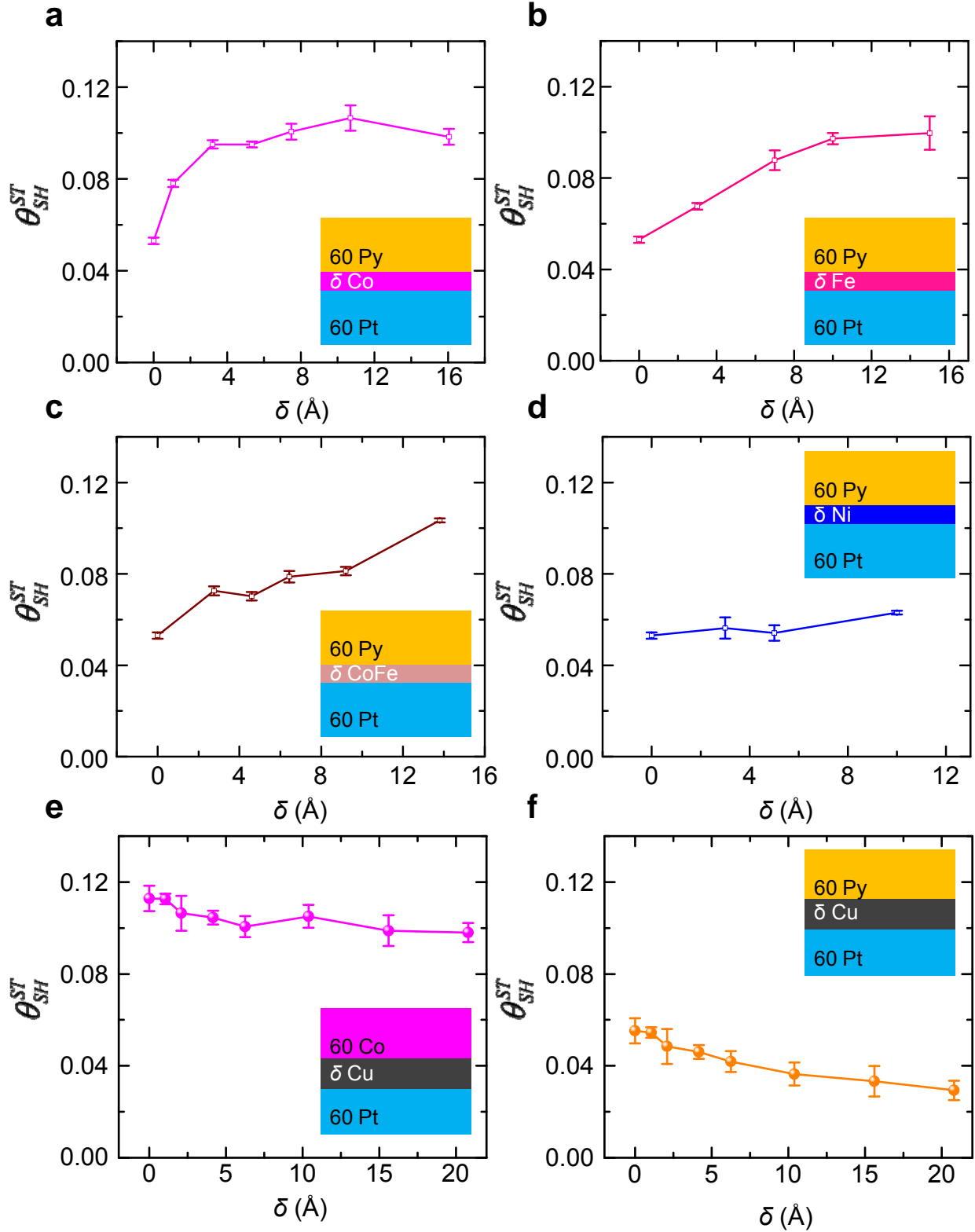